# Lotus: Creating Short Videos From Long Videos With Abstractive and Extractive Summarization

Aadit Barua*
Department of Computer Science
The University of Texas at Austin
Austin, Texas, USA
aaditbarua@utexas.edu

Karim Benharrak*
Department of Computer Science
The University of Texas at Austin
Austin, Texas, USA
karim@cs.utexas.edu

Meng Chen
Department of Computer Science
The University of Texas at Austin
Austin, Texas, USA
mengchen@utexas.edu

Mina Huh
Department of Computer Science
The University of Texas at Austin
Austin, Texas, USA
minahuh@cs.utexas.edu

Amy Pavel
Department of Computer Science
The University of Texas at Austin
Austin, Texas, USA
apavel@cs.utexas.edu

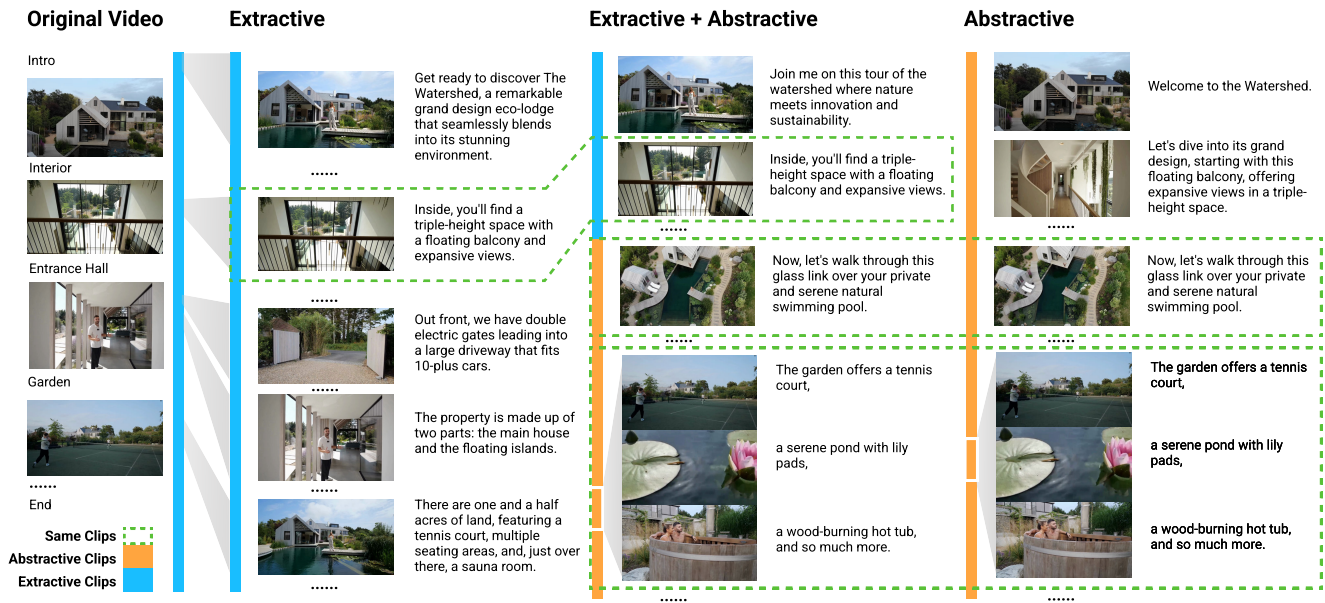

Figure 1: Lotus enables users to transform long-form videos into short-form by combining extractive and abstractive summarization. Compared to methods that use only extraction or abstraction, our approach balances the original connection between visuals and audio with the flexibility to cover more content.

## Abstract

Short-form videos are popular on platforms like TikTok and Instagram as they quickly capture viewers' attention. Many creators repurpose their long-form videos to produce short-form videos, but creators report that planning, extracting, and arranging clips from long-form videos is challenging. Currently, creators make *extractive* short-form videos composed of existing long-form video clips or *abstractive* short-form videos by adding newly recorded narration to visuals. While extractive videos maintain the original connection between audio and visuals, abstractive videos offer flexibility in selecting content to be included in a shorter time. We present Lotus, a system that combines both approaches to balance preserving the original content with flexibility over the content. Lotus first creates an abstractive short-form video by generating both a short-form script and its corresponding speech, then matching long-form video clips to the generated narration. Creators can then add extractive clips with an automated method or Lotus's editing interface. Lotus's interface can be used to further refine the short-form video. We compare short-form videos generated by Lotus with those using an extractive baseline method. In our user study, we compare creating short-form videos using Lotus to participants' existing practice.

---

*Both authors contributed equally to this research.

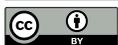

## 1 Introduction

In recent years, short-form videos have surged in popularity on platforms such as TikTok [14], Instagram Reels [10], and YouTube Shorts [22]. Short-form videos are fast-paced, engaging, and have the potential to reach a broad audience. Long-form video creators often repurpose their long videos into short-form videos. However, professional video editors in our formative study report that composing short-form videos from long-form videos is challenging and time-consuming. Creators must carefully review the video, select relevant clips, and then sequence and edit the clips together into a coherent story without introducing errors (*e.g.*, jump cuts in the visuals, or grammatical errors in the audio narration). When creators can not edit together existing clips to form a cohesive story, creators often re-script and re-record narration to make the video concise, coherent, and engaging. Creators in our formative study thus estimated that manually composing a single short-form video from a long-form video took between a few hours to a few days.

To support creators in converting long-form videos to short-form videos, commercial tools [2, 12] and prior research [42, 68, 71] let creators input a long video to receive short video options via *extractive* approaches that extract and edit together video segments. Prior work extracts clips by either considering the visuals or audio. *Visual-based extraction* approaches extract and play back important visual footage in a video (*i.e.,* a video skim) [42] but such approaches do not consider the audio and thus result in disjoint and nonsensical audio tracks. *Audio-based extractive* approaches such as Opus [12] and CapCut [2] extract a single coherent video clip from the long-form video to create a short-form video, while PodReels [71] and ROPE [68] extract then edit together multiple narration clips throughout the long-form video to create a coherent short-form video. PodReels [71] and ROPE [68] target audio or video podcasts such that they prioritize extracting a story with coherent and engaging narration. Such audio-based *extraction* preserves the original connection between audio and visuals, but limits the flexibility and content coverage of the output short-form videos. The *audio-based* approach also misses an opportunity to use rich visual content as b-roll (*e.g.*, for tours, reviews, travel, tutorials). We explore the potential to first create an *abstractive* short-form video and then add *extractive* clips to maximize flexibility, content coverage, and the original audio and visual connection.

We introduce Lotus, a system to support creators in combining *extracted* video clips and *abstractive* video narration together to create controllable short-form videos. Creators first provide Lotus with a long-form video and, optionally, specify preferred clips to guide the initial short-form video generation. Our system creates an initial *abstractive* short-form video by generating a short-form transcript and its corresponding speech, then matches relevant video clips with the generated narration. Creators can use Lotus to augment the abstractive short-form video with *extracted* video clips using our automated approach or editing interface. The editing interface allows creators to further refine their videos by changing the narration (via automated or manual editing) and finding visuals (via search, suggestions, and alignment) for each video clip.

To evaluate Lotus' automatically generated short-form videos, we conducted a results evaluation comparing an extractive baseline, our abstractive method, and our method combining abstractive and extractive techniques. We also conducted a user study (n=8) comparing the use of Lotus with the existing video editing tools used by creators to create short-form videos. Our results evaluation demonstrates the benefit of flexibility between abstractive, extractive, and abstractive-extractive approaches with raters preferring different approaches for different types of videos. All participants reported they wanted to use Lotus in the future and successfully used Lotus to author short-form videos with abstractive and extractive clips.

Our work contributes:

- Strategies and challenges that professional video editors face when creating short-form videos from long-form videos.
- Lotus, a system for creating *abstractive* and *abstractive-extractive* short-form videos from long-form videos.
- A user study demonstrating how creators use Lotus to make short-form videos.

## 2 Background

Our work on combining extracted video clips and abstractive video narration to create controllable short-form videos builds on prior research in video editing tools and video summarization.

### 2.1 Video Editing Tools

Traditional video editing tools such as Adobe Premiere Pro [1], VEGAS Pro [15], and Final Cut Pro [7] use a timeline-based editing approach where users scrub through a timeline to navigate the video and place cuts. While timeline-based editing supports fine-grained control, low-level edit interactions such as selecting the precise clip and placing cuts can be tedious and inaccessible [36].

To make videos easier to edit, research and commercial tools have introduced transcript-based editing, allowing users to edit videos by manipulating the narration transcript [25, 27, 31, 36, 54, 60, 63]. Berthouzoz et al. propose a tool to interactively place cuts based on the semantic content within the video transcript [25]. Similarly, Leake et al. present RoughCut, a tool that matches dialogue-driven video recordings to each line in a transcript based on user-specified film-idioms [45]. Beyond editing the original source, researchers have also explored systems to add audio or visuals to the input video. For instance, QuickCut allows creators to easily add a narration voiceover to the video by aligning footage shots to the narration script [66]. To add engaging visuals to the video, B-Script recommends B-roll insertions based on the video transcript [35]. Beyond directly interacting with the timeline or the transcript, LAVE enables users to provide commands to an LLM agent which interprets users' objectives and executes low-level editing interactions [69]. However, LAVE does not provide the flexibility to decouple or modify the audio and visuals which limits users' ability to tailor the content to specific purposes.

While prior work [68, 70, 71] and industry solutions [2, 3, 12] have investigated creating short-form videos, most systems focus on automating extraction or require additional recording. For instance, Opus [12] and CapCut [2] extract a single highlight clip, while PodReels [71] extracts highlights from video podcasts with monotonous visuals. ReelFramer [70] supports storyboard creation for news articles but requires new recording to create short-form videos. ROPE [68] concatenates relevant audio segments but does not consider visual elements, which can make it less suitable for short-form video creation where important information can be conveyed through visuals.

Our work explores how users create short-form videos given more control and flexibility over the audio and visuals. We propose a video editing tool where users can combine extracted video clips with newly generated abstractive narration.

## 2.2 Video Summarization

Video summaries can spark interest and anticipation for the full content (*e.g.,* teasers, trailers) or help the audience skim through the key moments of the video in a short time span [52]. One approach to summarizing a video is to use text-based summarization techniques on the transcribed speech or visual content [24, 53, 61]. There are two main techniques in text-based summarization: extractive summarization and abstractive summarization. Extractive summarization selects important sentences directly from the source to form a summary [23]. Abstractive summarization creates a short representation of the original text by generating new phrases and sentences [49]. Using extractive summarization, ROPE generates an audio summary of podcasts that preserves coherence of extracted audio segments [68]. To make an accessible audio description of videos, Rescribe uses word-level sentence simplification [56] to fit audio description into the gaps of video content. Using abstractive summarization, Li et al. [48] proposed hierarchical summarization of long-form spoken dialogue. Similarly, Video Digests [55] provide abstractive summaries of lecture videos to enable easy skimming.

In addition to text summaries, another video summarization approach is to generate a visual summary which is either a sequence of keyframes or moving images, called video skims [67]. Several video summarization datasets have been created that feature keyframes annotated with importance scores. For instance, SumMe [33] and TVSum [64] are datasets with video summaries composed of a subset of frames of the original video. Prior work [32, 40, 41, 73] evaluates models that predict the subset of the most important frames within an original video using benchmarks such as SumMe and TVSum. To give users control over extractive summarization, ElasticPlay [42] allows users to specify a time budget for the remaining content. While ElasticPlay enables users to determine the length of the summary, it does not allow them to select specific frames to include in the summary.

Existing approaches create either abstractive or extractive video summaries. We propose a method that combines these summarization techniques to create short-form videos, balancing the use of original content with more flexibility.

## 3 Formative Work

To inform the design of Lotus, we conducted interviews with 5 professional video editors and an analysis of 20 pairs of long-form and corresponding short-form videos from YouTube.

### 3.1 Short-Form Video Creation Study

**Participants.** We recruited 5 professional video editors (E1-E5) through Upwork. All video editors had experience creating short-form videos, with an average of 11.6 years (SD = 7.7) of video editing experience. All editors had experience transforming long-form into short-form videos across a range of video types (*e.g.*, short-form explainers, animations, commercials, and documentaries).

**Procedure.** We conducted a 1.5-hour study via Zoom that consisted of: (1) an interview to learn about their existing long-form to short-form video editing practices and (2) a 40-minute short-form video editing task. We started by asking video editors questions on their general video editing experience (*e.g.*, how many years of video editing experience do you have?). We then observed the video editors perform a video editing task to transform a long-form YouTube video into a short-form video for social media. They used their preferred software for the editing task (*e.g.*, Adobe Premiere Pro and Adobe After Effects) and chose from 5 long-form videos with an average video length of 12.13 minutes (SD = 3.35). Participants shared their screens and we invited them to think aloud during the task. We then conducted a post-task interview with questions reflecting on their editing process. We recorded, transcribed, and analyzed the study sessions using thematic coding to identify common themes. Participants were compensated at their hourly Upwork rate. This study was approved by our institution's IRB.

**How do editors create short-form videos from long-form?** Our observations with professional video editors revealed a consistent workflow for creating short-form videos from long-form videos.

Professionals first **determine a goal** for their video then carefully watch the long-form video with their goal in mind to **identify key clips**. Professionals reported that they derived their editing goals based on their own prior experience and client communication. E4 noted that they asked their clients *"what's your goal... How can I help you?"* and that *"some clients even write a book"* outlining their creative vision. Professionals then identified relevant clips by watching the entire video first to *"get a better understanding of what's going on"* (E4). E2 described identifying potential moments as going *"through the video"* to *"find the best moments."* All of the editors reported that they prioritize visually engaging content to capture and maintain audience interest. E5 cited the need to *"grab people's attention"* through visuals. Editors also reported that they considered additional factors such as clips that were requested by the client, clips that could transition well with other clips, clips that would serve as B-roll for narration, clips similar to effective clips in reference videos, and clips with catchy music. As editors selected clips, they considered whether each new clip would be coherent with their existing short-form video narrative. E2 emphasized that the *"message needs to be clear"* to the audience. All editors except E5 (who used CapCut [2] to automatically select a relevant clip) reported identifying clips to be challenging and time-consuming.

Editors must **arrange selected clips** into a coherent sequence. E2 characterized this step as finding *"some kind of flow that works for those shorter moments."* E4 highlighted that trimming and arranging video clips is particularly challenging and usually takes the longest time. Editors then **add external elements** such as narration, sound effects, text overlays, and graphics that were not in the original video to make their short-form video more engaging and align it with their higher-level goals. Editors adjust the aspect ratio, typically to 9:16 dimensions, to meet platform requirements. Editors finally **create a rough cut** for feedback from clients or themselves that they can use for iterative improvements. E2 mentioned they create *"a rough edit"*, then seek input to *"get some eyes on it to give feedback"* and evaluate its alignment with the higher-level goals.

While video editors follow a structured workflow, they must balance technical skills, creative decision making, and client communication. Editors reported that crafting a short-form video from a long-form typically took them: 2-3 days (E3), 6-20 hours (E2), 2-3 hours (E1), 1-2 hours (E4), and 5-30 minutes (E5, used automated single-clip extraction via CapCut [2] rather than editing).

**What makes a good short-form video?** Video editors shared insights on what makes a good short-form video including a clear concise message (E2), engaging visuals (E1), and a compelling hook within the first 3 seconds (E4). Video editors also mentioned that graphics, transitions, and text overlays are an effective way to make the video more engaging and understandable.

**Future and prior use of AI-powered tools.** E5 (the most experienced participant in AI tool usage) mentioned that Opus offers *"very limited"* editing options after generating automatic short-form videos and expressed their desire for *"more... editing options for Opus."* Looking ahead, E1 envisioned a system that would perform *"intelligent work"* to create a shorter version from which they could *"take ideas from"* and *"come up with their own version."* E4 emphasized that *"you always need a human... to really be creative"* and put *"their own spin into it,"* but they expressed interest in *"having like a partner"* that makes the workflow *"faster, easier, [and] simple."*

### 3.2 Analysis of Short-form Videos

**Procedure.** We collected 20 long-form and short-form video pairs on YouTube by searching across a diverse range of creators and content categories. Our sample included creators who published corresponding short-form versions (a YouTube Short) of their long-form content. Long-form videos averaged 17.9 minutes (SD = 7.9), while short-form versions averaged 47.5 seconds (SD = 19.2).

**Short-Form Video Types.** Videos fell into two distinct categories: *highlights* that extracted an engaging or memorable moment (13 of 20) and *summaries* that edited together multiple clips throughout the video (7 of 20). All short-form videos directed viewers to the full-length version: 18 of 20 included a link in the description and 2 of 20 included explicit verbal calls to action.

**Visuals.** 15 of 20 short-form videos reused visuals from their long-form counterparts. Some videos used entirely new or previously unused footage, such as alternate camera angles or content potentially shot for the short-form version. We also observed split-screen layouts to show multiple perspectives of the long-form visuals during narration. 13 of 20 short-form videos contained captions (typically of the narration).

**Audio.** 13 of 20 short-form videos added new audio elements, **primarily new narration over reused visuals**, to maintain narrative coherence in the shorter format. Another strategy involved adding background music to enhance engagement. When reusing the long-form audio, several videos decoupled the audio and visuals by matching narration from one part of the long-form video with visuals from another.

### 3.3 Design Goals

Based on our formative study with professional video editors and our video analysis of long-form and short-form video pairs, we identified three design goals for our system.

**G1** *Identifying Key Clips*
Video editors must identify clips that are visually engaging, contain important information, and convey a coherent narrative. Our system should help users identify key clips within the video by considering the visuals and audio.

**G2** *Arranging Clips*
Video editors must arrange selected clips into a coherent sequence. Our system should support users arranging clips to create a coherent and engaging short-form video.

**G3** *Flexible Audio and Visuals*
In existing short-form videos, editors frequently reuse existing video clips and replace the audio track with new narration to create a coherent narrative. Our system should help users create a short-form video unrestricted by the original audio and visual synchronization.

## 4 Lotus

Lotus supports the creation of short-form videos by enabling users to rewrite narration (abstractive clips) and extract original content (extractive clips). To combine abstractive and extractive clips, Lotus uses a two-stage video generation pipeline with an editing interface for further refinement. We begin with an example workflow for creating short-form videos using Lotus, followed by an overview of the features in Lotus's editing interface. Finally, we describe Lotus' pipeline. Lotus is implemented using a React frontend, a Flask backend, and a separate Flask API for user interaction logging.

### 4.1 Editing Workflow

To illustrate how Lotus works, we share an example scenario following Alice, a video editor who works for a real estate agent client. Alice needs to transform a 15-minute YouTube house tour video into a 1-minute short video that her client can publish to increase the exposure of the house on sale.

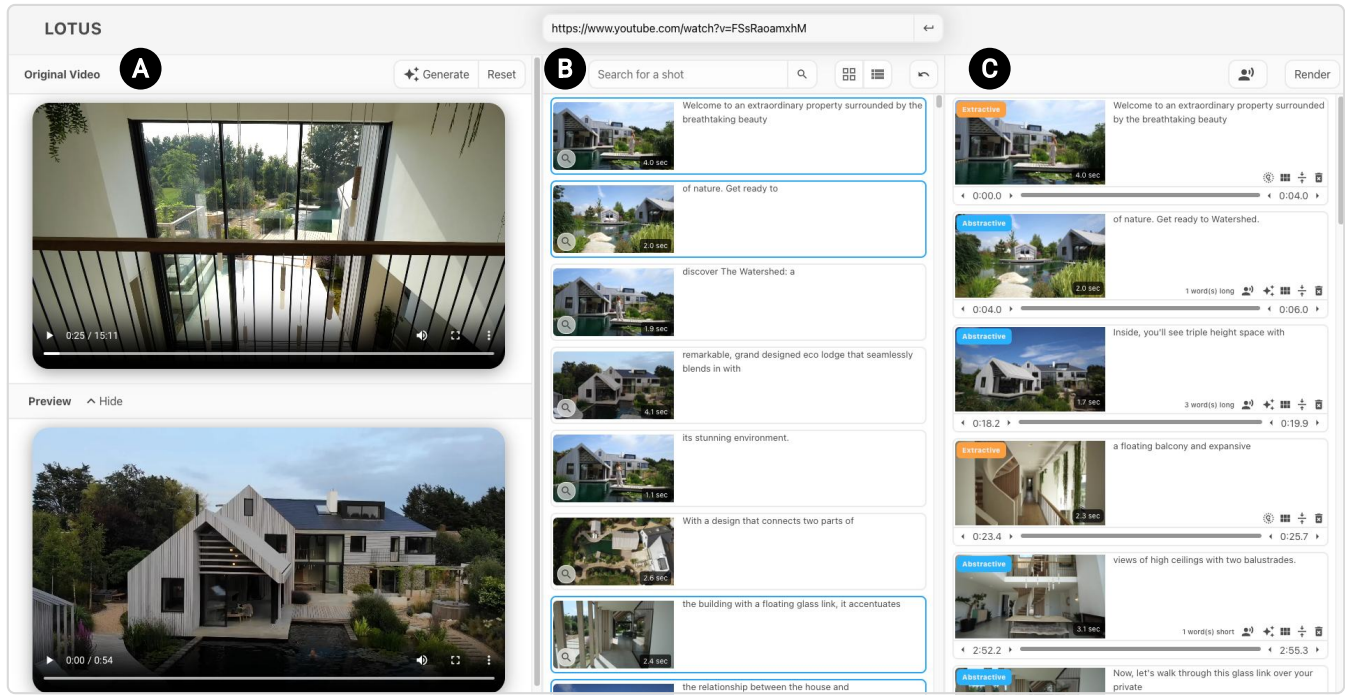

**Figure 2: Lotus interface consists of 3 panes: (A) Video Player Pane, (B) Long-Form Video Clips Pane, and (C) Short-Form Video Clips Pane.**

Alice opens Lotus and inputs a long-form video. Lotus processes the video and presents Alice with clips from the long-form video (Fig. 2B). Skimming the keyframe and speech of each clip, Alice plans out her short-form video and selects clips to guide the generation process (*e.g.*, staircase, kitchen, swimming pool, bedroom). Alice clicks the generate button, prompting Lotus to create a short-form video and populate its clips in the right-most pane (Fig. 2C).

Alice goes through clips in the generated short-form video and examines whether they match the plan she made. She doesn't like the speech for one clip because it calls the pond a pool. She rewrites this speech and chooses to also highlight the pond's natural surroundings. Noticing that the video features only a partial close-up of the staircase while the narrator is introducing its spiral shape, Alice searches *staircase* to find a clip that that captures the entire staircase. Alice also finds the initial list of clips includes a lily pad which is a minor detail of the house and decides to delete it. Alice decides to look for a conclusion. She likes the way the long-form video concluded so decided to extractively bring the concluding clips into the short-form video. Lastly, Alice trims all short-form clips to avoid jump cuts.

Once Alice has finalized her editing of clips, she clicks the render button and previews the entire short-form video in the video player. She proceeds to send the 1-minute video from Lotus to her client.

### 4.2 Interface and Interactions

Lotus has 3 main panes that users edit and evaluate the videos in: **Video Player Pane** (Fig. 2A), **Long-Form Video Clips Pane** (Fig. 2B), and **Short-Form Video Clips Pane** (Fig. 2C).

*4.2.1 Video Player Pane.* This pane consists consists of the Original Video Player (Fig. 2A, top) and the Preview Player (Fig. 2A, bottom). In the Original Video Player, users can watch the entire long-form video or play individual clips by clicking on keyframes from either the Long-Form or Short-Form Video Clips Panes. The Preview Player allows users to view the latest version of the short-form video, which updates each time the render button is pressed.

*4.2.2 Long-Form Video Clips Pane.* The Long-Form Video Clips Pane allows users to navigate the long-form video content at the clip level in either a grid view (showing only keyframes) or a list view (displaying both keyframes and speech). Users can perform a CLIP-based [59] search by entering a prompt or selecting a clip's keyframe, which reorders clips based on the search results **(G1)**. To restore the original order of clips, users can click the revert button.

*4.2.3 Short-Form Video Clips Pane.* The Short-Form Video Clips Pane presents a list of the current short-form video clips, each displayed with its keyframe and associated speech. Users can manage clips by dragging, dropping, and deleting them between the Long-Form and Short-Form Video Clips Panes. To support users creating short-form videos, Lotus offers the following features:

- **Toggle:** Users can toggle between abstractive and extractive modes for any clip.

- **Speech Editing (G3):** In abstractive mode, users can edit the text for a clip's speech.
- **Generate Alternative Text (G3):** In abstractive mode, users can choose to generate new text for a clip's speech (see 9.1).
- **Speaker Selection (G3):** In abstractive mode, users select from narrators in the video (found using PyAnnote speech diarization [58]) or a default voice for speech generation.
- **Denoise:** In extractive mode, users can remove background noise to maintain audio consistency across clips.
- **Trim and Extend:** Users can trim and extend clips to their desired length based on the corresponding speech length.
- **Alternative Clips (G1):** Users can replace clips with alternative clips when they prefer different visuals. (Fig. 3).
- **Align (G2):** Users can align short-form clips with their position in the Long-Form Video Clips Pane to see the surrounding context.

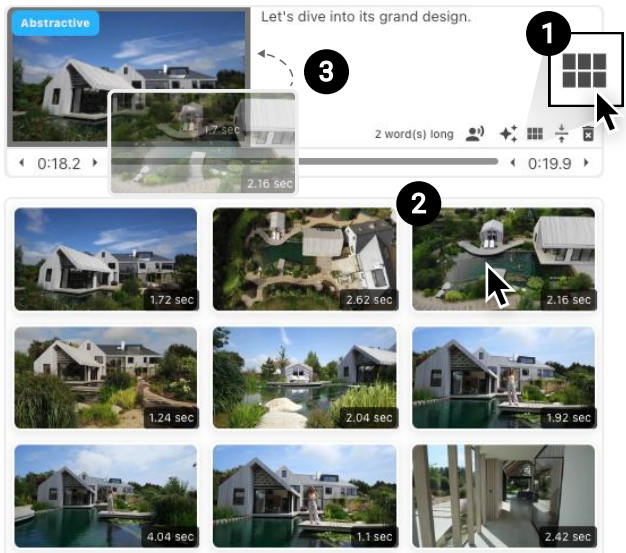

**Figure 3: Users can choose from alternative clips if they prefer different visuals for a specific transcript section. Users first click a button (1), then choose from a set of 20 alternative clips that match with the transcript section (2). Clicking on a clip replaces the prior clip (3).**

## 4.3 Pipeline

Lotus's pipeline creates an initial short-form video composed of abstractive clips and then extends the video by automatically adding extractive clips. Audio and video processing are handled using Moviepy [11] and torchaudio [37].

*4.3.1 Short-Form Transcript Generation.* Lotus first generates a transcript that is a summarized version of the long-form transcript by leveraging GPT-4o [8]. We construct the prompt such that the generated text maintains the original style, is suited for social media platforms, and has a strong hook (see 9.1 Prompts). To ensure our generated short-form transcript captures important visual content, we extract the noun phrases (using spaCy [34]) in the long-form transcript and include them in the prompt **(G3)**.

*4.3.2 Visual Concepts.* Similar to prior work [47, 72], we identify visual concepts (*i.e.*, words or phrases with clear visual connection) within the short-form transcript **(G1)**. First, we run scene detection [28] to extract clips from the long-form video. Next, we provide GPT-4o [8] with visual descriptions of these long-form clips (generated by GPT-4o) along with the short-form transcript to extract the relevant visual concepts (see 9.1).

*4.3.3 Clip-Transcript Matching.* Lotus aligns long-form clips with the short-form transcript using the identified visual concepts and a weighted scoring function. We score and assign each long-form clip to a corresponding visual concept in the short-form video transcript using four metrics **(G1, G2)**:

- **Speech Similarity:** We measure the similarity between a long-form clip's transcribed speech and the target visual concept, using nomic-embed-text embeddings [51].
- **Keyframe Similarity:** We use CLIP [59] to calculate the similarity between a long-form clip's keyframe and the target visual concept.
- **GPT-4o Scoring:** We prompt GPT-4o [8] to score approximately 25 clips (filtered using the 3 other scoring metrics) based on similarity to a target visual concept. We provide each clip's keyframe and speech as context (see 9.1 Prompts).
- **Position-Based Alignment:** We score how closely the relative position of a target visual concept aligns with the relative position of a clip in the long-form video. The distance between these positions is calculated as follows:

$$Score_{pos} = 1 - Pos(\text{visual concept}) - Pos(\text{long-form clip})$$

We assign the highest-scoring long-form clip to each visual concept in the short-form video transcript. To eliminate duplicate clips, we iteratively replace them with the next highest-scoring alternative until no duplicates remain. To create the initial abstractive video, we generate speech for the short-form transcript using ElevenLabs [5] and align this generated speech with long-form clips matched to visual concepts. To align the duration of each clip with the generated speech, we adjust clip playback speed as needed. This abstractive video serves as the initial generated result presented to users in our interface **(G2)** and forms the basis for the second phase of our pipeline, which blends abstractive and extractive clips.

*4.3.4 Blending Abstractive and Extractive Clips.* To create a short-form video that preserves the original audio-visual connection while enabling flexibility over content, the last part of our pipeline combines abstractive clips from the initial video with extractive clips from the long-form video.

First, both the transcript from our initial abstractive video and the long-form video are segmented. We segment the initial abstractive video transcript using GPT-4o (see 9.1), while the long-form video transcript is segmented using ROPE's algorithm [68]. We score and compare each abstractive segment against each extractive segment based on the following four metrics **(G1, G2)**:

- **Speech Similarity:** The similarity between an abstractive and extractive segment's speech (see 4.3.3).

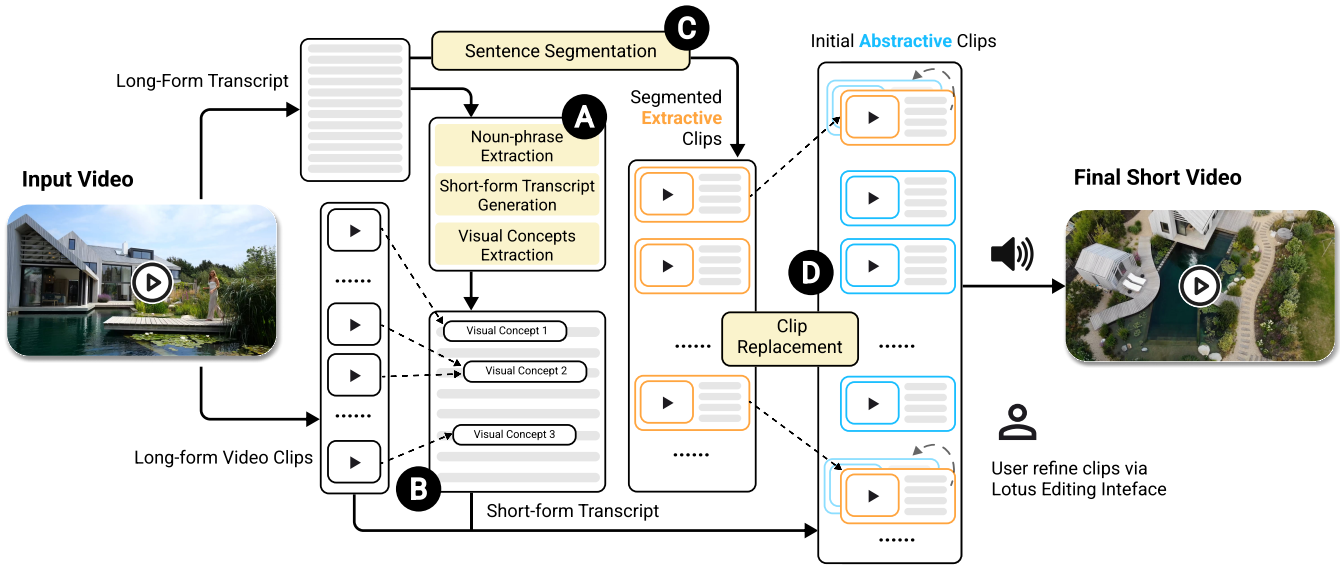

Figure 4: Lotus short-form video generation pipeline. Lotus first segments the input video into clips and transcribes the speech into text. Using the long-form transcript and extracted noun phrases, Lotus generates a short-form transcript and identifies key visual concepts (A). A weighted scoring function maps the long-form video clips to the visual concepts, producing an initial list of abstractive clips (B). Extractive clips are created from sentence segmentation similar to ROPE [68], based on the long-form transcript (C). Lotus replaces selected abstractive clips with extractive ones using a clip replacement scoring function (D).

- **Visual Connection:** The similarity between a segment's speech and its keyframe, using CLIP [59].
- **Coverage:** The number of noun phrases occuring within a segment's speech, found using spaCy [34].
- **Position-Based Alignment:** The distance between the relative position of an abstractive and extractive segment.

If the difference between an abstractive segment and the highest-scoring extractive segment falls below a threshold (empirically determined), we create permutations that allow each option as a possibility. From this set of permutations, we select the one with the highest coherence score, determined by GPT-3 [26] loss. To produce the final short-form video, we extract clips from each video based on the chosen segments and sequence them together. We then apply volume normalization [6] and denoising [13] for consistent sound quality across both abstractive and extractive clips.

## 5 Results Evaluation

We evaluate the short-form videos produced by our pipeline, by comparing our abstractive and blended abstractive-extractive (also referred to as mixed) methods with an existing extractive audio-shortening method [68].

**Method.** We first selected six videos from YouTube's trending section that met the following criteria: between 5 to 20 minutes in length with visually diverse content (*e.g.*, not podcasts, TED talks, or lectures) Detailed information about the selected videos is provided in Appendix Tableg 2. We processed each video by generating results for Lotus's two methods and the extractive baseline. The extractive method described in [68] was designed for audio shortening, so we adapted it for short-form video creation. We re-implemented the sentence segmentation, sentence score/length calculation, and combinatorial optimization as described. For the abstractive summarization, we replaced BART with GPT-4o [8]. The long-form video transcript is used as input, providing selected sentences as output. We then create the short-form video by extracting and sequencing video clips based on the timestamps of these selected sentences.

Participants rated examples (3 methods per example) and we qualitatively evaluated the results to identify common errors. We recruited the 12 annotators through university mailing lists who have experience watching short-form videos. Participants ranked the short-form videos produced by each method by overall preference (Figure 5). Each video was also rated based on its hook, narrative, and visual appeal using a 5-point Likert scale (Figure 9).

**Results.** Overall, annotators rated the blended abstractive-extractive (mixed) results ($\mu$ = 1.88, $\sigma$ = 0.72) as more preferred than the abstractive ($\mu$ = 2.06, $\sigma$ = 0.77) and the extractive method ($\mu$ = 2.04, $\sigma$ = 0.82), but this difference was not significant. While aggregate preferences were similar across conditions, participants had method preferences for different videos due to the content of the video and the quality of the result (Figure 5).

Annotators preferred extractive and mixed approaches over the abstractive approach for videos in which the **narrator was prominently speaking to the camera** (V2 and V3). In these videos, while the abstractive result achieved high coverage of the video content (*e.g.*, all the steps of a cooking video for V2), the synthesized voice with illustrative visuals in the abstractive approach can feel less engaging than a speaker looking at the camera and talking (only possible with extractive and mixed). However, the abstractive approach was preferred when the **narrator was not prominently featured** in the long-form video or extracted clips (V4 and V5).

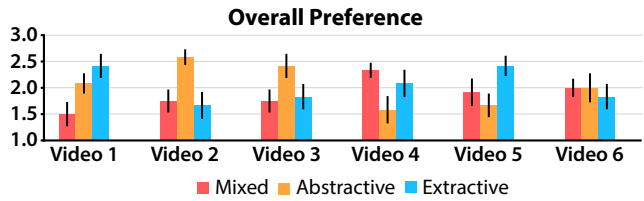

Figure 5: Average ranking of overall preferences within the results evaluation grouped by video for the mixed abstractive-extractive, abstractive, and extractive methods. A lower rating means that it was preferred more. Error bars represent a 95% confidence interval.

The **lack of a clear introduction** (or "hook") or a **missing conclusion** appears as an obvious error and can impact preference. For example, the extractive version of V5 was missing a hook and V1 was missing a hook and conclusion — annotators thus rated the extractive version lowest for these videos (Figure 5). An existing fast-paced hook or conclusion sometimes does not exist in the long-form video, or the extractive approach misses it. In V1, the mixed video primarily featured the abstractive storyline but also replaced the introduction of the video with an extracted hook, and annotators preferred the mixed approach most often.

Overall, both abstractive and extractive selections provide different advantages, prompting the need for flexibility in choosing between both when creating short-form videos. Lotus' interface thus lets users flexibly combine abstractive and extractive selections. We take advantage of the strength of the initial abstractive storyline to create an initial result, then let users flexibly add extractive footage to increase engagement and realism.

## 6 User Study

To understand how Lotus supports users in their video editing workflow for creating short-form videos, we conducted a within-subjects study with 8 participants. To evaluate our system against current practices, we established a baseline consisting of participants' preferred video editing tool (*e.g.,* Adobe Premiere Pro). Each participant was provided with two long-form videos and instructed to create two short-form videos, one per method. Participants completed a questionnaire on result satisfaction, task load, and creativity support after each method. Finally, we conducted a semi-structured interview to gather further feedback on their experiences. All interviews were recorded, transcribed, and analyzed using thematic coding by two researchers to identify common themes. Our study investigated the following research questions:

**RQ1.** *How helpful is the initial short-form video generated by Lotus?*
**RQ2.** *How does Lotus support video editors in creating short-form videos from long-form videos?.*
**RQ3.** *How does Lotus and short-form videos created using Lotus compare to video editor's traditional approaches?*

### 6.1 Participants

We recruited 8 participants with video editing experience through Upwork and university mailing lists. Participants were compensated according to their Upwork rates or $50. Participants had and average age of 22 years (SD = 3.46) and reported an average of 3.71 years (SD = 4.65) of video editing experience. Participants previously created videos for social media platforms like YouTube (P1, P3, P4, P6, P7), school projects (P1, P2, P3, P5), and Television and Film (P6) with video types ranging from ads (P5, P6) to instructional videos (P1, P5, P7), vlogs (P2, P5), and montages (P4). Two participants reported having little experience with AI tools for video editing (P2, P6), while all other participants had no experience.

### 6.2 Procedure

For our study, we selected two long-form videos from YouTube's trending section based on the following criteria: have similar content and length, be between 10 and 20 minutes in length, and have visual diversity (excludes videos like podcasts, music videos, talks). The selected videos had durations of 13 minutes and 57 seconds and 15 minutes and 11 seconds. The study was conducted both in person and remotely over Zoom. To conduct the study remotely, participants were given access to our system through Zoom's Remote Control features.

To familiarize participants with our system, they were provided with a short tutorial video and interactive section before beginning the editing task. Participants were randomly assigned to conditions with a counterbalanced order to prevent a learning effect. Each participant had a maximum of 20 minutes to complete each task. Before each editing task, participants were given 10 minutes to familiarize themselves with the long-form video they were going to edit. The entire study lasted approximately 1 hour 45 minutes.

### 6.3 Results

All participants stated that they would like to use Lotus in the future, as it was *"easy to use"* (P3). P6 said that they *"spent years of their life looking for another software that does half of what [Lotus] does."* Participants created short-form videos with Lotus that they found comparable in quality to those made with their existing tools (Figure 7) without experiencing increased mental demand during the editing process (Figure 6). However, participants liked the editing process within Lotus over their existing tools (Figure 7). P2 appreciated how *"Lotus gives a summary of the entire video hitting every point so what took 20 minutes with Lotus would have taken 40 minutes with CapCut."* P2 *"really liked how [Lotus] generated a base to go off of."* All participants refined the initial generated short-form video using abstractive and extractive clips (Table 1) compared to the fully extractive videos for the baseline.

*6.3.1 Editing Workflow.* All participants generally followed a similar workflow when using Lotus. Participants described how Lotus gave them a fresh start when creating short-form videos while their existing tools started with the entire long-form video which they had to cut down from. P8 said that Lotus *"gives you more freedom because it feels like you're going from scratch"* compared to existing tools where *"it's really hard to just cut out everything."*

| ID | Align | Edit Text | Drag Shot | Delete Shot | Trim Shot | Render |
|---|---|---|---|---|---|---|
| P1 | 8 | 48 | 24 | 6 | 65 | 6 |
| P2 | 5 | 19 | 15 | 5 | 58 | 6 |
| P3 | 0 | 11 | 15 | 11 | 77 | 3 |
| P4 | 2 | 5 | 35 | 19 | 66 | 3 |
| P5 | 4 | 25 | 40 | 14 | 39 | 3 |
| P6 | 0 | 1 | 8 | 2 | 25 | 1 |
| P7 | 4 | 33 | 21 | 4 | 157 | 4 |
| P8 | 4 | 3 | 65 | 7 | 76 | 1 |

**Table 1: Participant interactions with Lotus. Participants used *align* to align the current short-form video with the long-form video, *drag shot* to add long-form shots to the short-form video, *delete* and *trim shot* to remove or trim shots in the short-form video, and *render* to preview a revised result.**

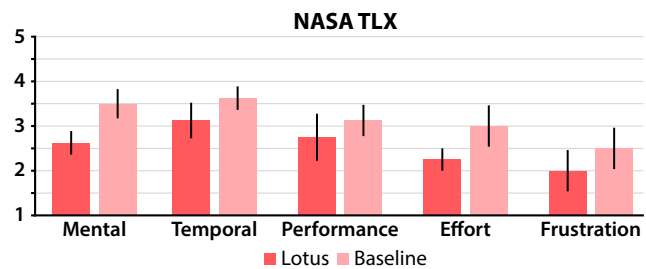

**Figure 6: Participant ratings for NASA TLX. Error bars represent a 95% confidence interval.**

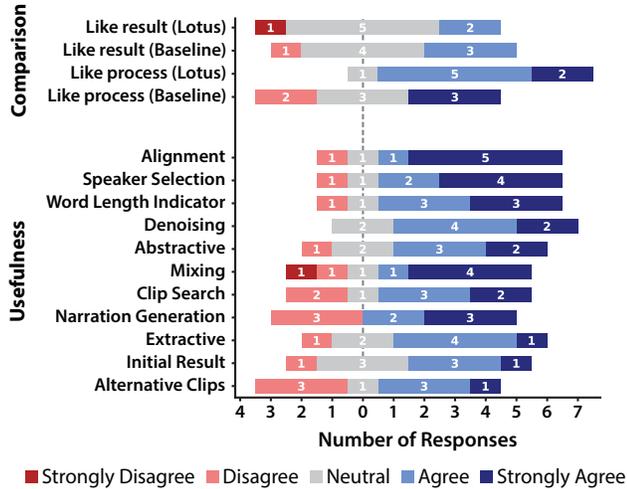

**Figure 7: Subjective participant ratings on a scale from 1 (strongly disagree) to 5 (strongly agree).**

**Generated result serving as initial structure.** All participants found value in the initial result generated by Lotus. P2 noted that it matched their mental mapping of the content: *"I made a mental note of which clips I wanted to use and Lotus chose those ones for me already."* All participants started by reviewing the entire initial result. Participants (P1-P3, P5, P7, P8) saw the initial result as a foundation for their editing process which provides structure and inspiration for further refinement. P2 said they *"they really liked how it generated a base to go off of because after watching the original video you kind of forget the main clips."* Some participants (P4, P6) focused less on the structure provided by the initial result and started their process by going straight into the Long-Form Video Clips Pane to identify clips to include in their short-form video.

**Adding content to the short-form video.** After watching and skimming the initial result, participants looking to add new clips turned their attention to the Long-Form Video Clips Pane. Instead of directly scrolling through the pane, we observed participants using the generated result as a summary as they aligned specific short-form clips. This enabled quick navigation and skimming of the relevant areas in the long-form video to their current short-form video. Participants found the align feature useful for finding surrounding context to refine transcripts and add nearby clips. P2 said *"the align feature let me go back and take clips or make a more coherent sentence."* Participants cited that they added clips to include more details, storytelling moments, *"attention grabbers,"* and improve visual aesthetics.

During editing, all participants focused on making sure their short-form video clips covered the details they wanted to convey to viewers. P1 said when it came to adding clips, they were *"more focused on the aesthetic of it."* According to P2, the initial result lacked some storytelling clips and they wanted to include these *"more emotional parts."* P8 described that *"for viewers just scrolling through Instagram or TikTok what they want to see is attention grabbers."* They said a *"wide shot of water crashing into the house that's a very good attention grabber."* To include *"attention grabbers,"* we observed participants select extractive clips with background music and no narration. To add content that was not yet present in their current short-form video, participants often had a specific clip in mind. For these situations, they found the prompt-based search useful for identifying clips. P6 envisioned its use in working with a wedding video where they want pick specific clips of the bride and groom: *"You have a wedding everybody walking down the aisle then you have a bride on stage. However, you want to see whenever they're on stage. You search the wording, grab the clip and done."*

**Removing content from the short-form video.** For removing clips from their short-form video, participants removed unnatural-sounding or long speech and unappealing or irrelevant visuals. Participants reported that the speech of clips generated by AI sometimes *"sounded funny"* (P4) or *"did not give enough personality"* (P1). Both abstractive clips in the initial result or extractive clips could be excluded by participants because they were too long or not engaging. P8 described excluding clips with *"too much of the unnecessary details"* since viewers would not find it engaging. Similar to adding clips for content coverage and aesthetics, clips were removed for being visually unappealing or irrelevant to the short-form story.

**Editing clips to refine narration.** When participants were not satisfied with the speech of specific clips within the short-form video, they often edited it instead of deleting the clip. Most commonly participants were making small edits like removing a few words to improve coherence and change word spelling. We observed participants use the align feature to get inspiration from the related long-form video clips to write speech. However, sometimes participants chose to alter speech through their own experimentation. Some participants found this difficult and used the text generate feature instead. P1 described the feature as good for *"brainstorming and finding a jumping off point since it can be hard to come up with the first idea."* For their abstractive clips, the majority of participants used a single speaker but a few (P5, P7) used multiple speakers to improve their short-form narrative. All participants used the trimming feature to refine the narration (for extractive clips) and cut transitions between clips in their short-form video.

**Evaluating the final result.** After 20 minutes were reached or participants were satisfied with their short-form video, they reviewed their result. Most participants (P1-P5, P7, P8) were happy with the short-form video they created using Lotus. When asked what changes participants would apply to their created short-form video provided with more time, the most common feedback (P3, P5, P6, P7, P8) was refining the transitions to avoid abrupt cuts.

*6.3.2 Abstractive vs. Extractive Clips.* Participants showed preferences for both abstractive and extractive clips with each having their respective benefits. Abstractive clips were favored for their flexibility, allowing participants to *"say exactly what they needed to say"* (P3). This allowed them to use a clip that had visuals they liked even if the audio did not fit with where they intended it to go. Participants also found this useful when an extractive clip had long narration. Instead of deleting the clip they could make it abstractive and just summarize the speech. P8 said *"if something could be shortened they would use abstractive."* Participants generally thought extractive clips should be viewed as their default option. P1 described this as extractive clips having *"more credibility"* because *"you can tell it's somebody who authored the video talking about it."* The visuals and audio of extractive clips sound very natural because the author planned that out for the long-form video. Participants highlighted that the voice of extractive clips retained more personality compared to AI generated voices. They further mentioned that the extractive narration sounds less artificial than the AI generated narration. P7 mentioned that they would try to establish the narrator once in the beginning with an extractive clip so that following abstractive clips appear more natural.

*6.3.3 Lotus vs. Existing Approaches.* All participants enjoyed the editing experience with Lotus describing it is painless (P1) and easy to use (P3). They highlighted how Lotus allows them to achieve a better final product more easily compared to tools like CapCut (P2).

Participants mentioned how Lotus reduces their time and manual work. Some participants highlighted how they preferred Lotus as it provides a simpler interface compared to existing tools (e.g., iMovie [9], DaVinci Resolve [4]), which often have features that are difficult to find and learn. P3 described how iMovie is *"very sleek but you have to like hover and keep clicking around to find what you need."* Participants (P2-P4, P7, P8) highlighted the benefits of transcript-based editing which allows them to see the captions for each clip which they can use to make editing decisions.

However, participants noted several advantages of their existing tools, such as a more functional editing timeline, more granular trimming, and immediate feedback. Other advantages mentioned included disconnecting an audio track from it visuals (P4) and turning the audio track off (P6).

Participants found the timeline in their existing tools more functional, primarily because of their horizontal layout compared to Lotus's vertical layout. This layout allowed participants to navigate the long-form video more easily. Another feature of timelines in existing tools is the ability to perform more granular cutting and trimming, allowing users to sub-clip and adjust to any length. This compares to a fixed granularity for trimming in Lotus. While participants valued immediately seeing the effects of their changes before making further edits, users using Lotus did not experience immediate edit feedback due to speech generation latency. In the future, Lotus could be integrated into existing editing tools to leverage features like traditional timeline-based editing.

## 7 Discussion

Findings from our results evaluation and user study suggest that combining abstractive and extractive summarization techniques offers a promising approach to creating short-form videos from long-form content. We reflect on the development and evaluation of Lotus, opportunities for future work, and ethical considerations.

**Effectiveness Across Video Types**. Our results evaluation highlighted that Lotus' performance varies across different video genres. Videos with prominent narrators, such as a spotlight interview, benefit from extractive approaches, as the original alignment between audio and visuals enhances authenticity. Conversely, abstractive methods are preferred for videos with less visually tied narration, such as travel vlogs, as their flexibility enables more concise storytelling. Instructional videos posed unique challenges as they rely on step-by-step instructions and precise alignment between narration and visuals. Lotus' reliance on existing visuals made it difficult to visually include all the important instructions in a short-form video. Future work could explore solving this by adding B-roll recommendations (similar to B-Script [35]) or generative B-roll creation.

**User Preferences and Interaction**. Our user study revealed the value of the initial short-form video provided by Lotus. Participants found the generated result to be a good starting point and often retained many moments initially selected by Lotus. The editing interface enabled users to refine clips, add new ones, and adjust the narration, providing users with support and full control over their short-form video. Participants varied widely in how they used these tools: some made only minor edits to the initial results (*e.g.*, P8, P6, and P4 edited text fewer than five times), while others extensively modified the video, using features like alignment and speech editing to tailor their videos. Future iterations of Lotus could provide multiple starting options, such as extractive, abstractive, or blended methods, allowing users to choose the approach that aligned with their preferences. Additionally, a natural language interface (similar to LAVE [69]) could help users describe high-level goals or even make simple edits.

**Speech Generation Quality**. Participants noted that generated speech sometimes lacked the expressiveness and naturalness of recorded audio, which reduced the appeal of abstractive clips. Prior work describes the lack of paralinguistic voice features, such as appropriate intonation and emotional expression, as a reason for perceiving synthesized voices as unnatural [38]. The unnaturalness was amplified when creating short-form videos featuring multiple speakers, as users could more easily perceive artifacts across different voices and generations. Supporting editors with fine-grained emotional control over the synthesized voice may improve its expressiveness and perceived naturalness [29, 39, 43]. Additionally, speech editing models can be used to remove artifacts within generated speech, for example, at speaker boundaries [57].

*Compelling Hooks and Conclusions*. Our evaluation revealed the outsized impact of certain key moments, such as hooks, conclusions, and on-screen narrators, on video quality. Missing or weak hooks diminished viewer engagement (Figure 5). While Lotus does not surface specific key moments, Lotus' interface supports users to quickly skim and extract moments, ensuring narrative coherence and engagement. However, automating the detection or generation of hooks and conclusions remains an area for future exploration. Possible approaches include asking the editor to provide explicit clues that can be factored in within the scoring function (similar to PodReels [71]), including a new metric in the scoring function that measures the uniqueness of frames from all other frames occurring in the original video, or using a more user-based approach where Lotus' algorithm can be used to automatically convert long-form videos into short-form videos based on viewers' interests, for instance, derived from their prior watching history [30, 44].

*Integration With Existing Tools*. Lotus is not as fully functional as traditional video editing tools like Adobe Premiere Pro. does not provide a comprehensive suite of features that can be found in traditional video editing tools such as Adobe Premiere Pro [1]. These features include timeline-based editing, changing the aspect ratio, and separate editing of audio and video channels. Consequently, participants expressed their wish for features that allow them to set a horizontal layout and have more granular trimming options. Future iterations of Lotus can draw inspiration from video editing tools presented by prior work to support these interactions [65, 66]. To support editors in making more fine-grained edits during their last iterations before publishing their material, Lotus can be extended with a way for editors to export their edited material for import into their traditional editing tools, similar to ChunkyEdit [46].

*Ethical Considerations*. The AI-powered capabilities of Lotus create ethical concerns around content attribution and misrepresentation. Lotus supports the rapid creation of short-form videos from long-form content, raising concerns of misuse, such as reposting content without proper credit to the original creator. The ability to rewrite and generate narration in Lotus introduces the potential for altering or misrepresenting the original message, particularly when repurposing other creators' content [50, 62]. Future systems should implement mechanisms, such as labeling modified clips, to enforce content attribution and prevent misuse.

## 8 Conclusion

In this paper, we present Lotus, a system that supports the creation of short-form videos from long-form content by combining the strengths of abstractive and extractive summarization. Our formative study identified a consistent workflow, common challenges, qualities of a good short-form video, and the role of AI in the editing process. From these insights, we created design goals that shaped the development of key Lotus features. Our results evaluation of Lotus suggests that participants' method preferences varied depending on the composition of the long-form video, such as if it featured a talking head. In our user study, participants successfully used Lotus to create short-form videos comparable to those produced with traditional tools, finding the editing process intuitive and enjoyable. Participants valued Lotus' initial video as a starting point for refinement, aligning short-form clips for added context, and the flexibility to switch between extractive and abstractive modes. Our findings highlight the potential of Lotus to support video editing workflows and the creative demands of short-form video creation.

## 9 Appendix

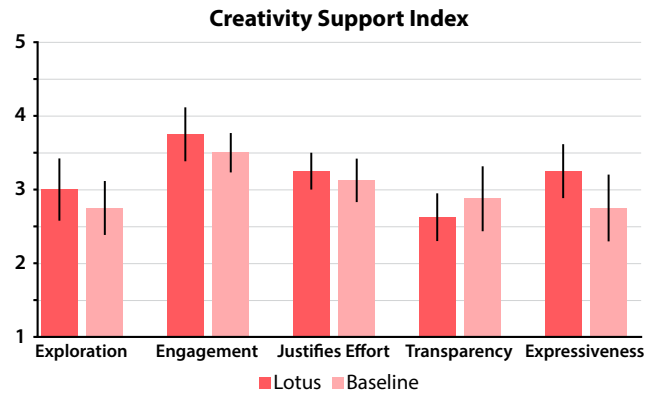

Figure 8: Participant ratings for Creativity Support Index. Error bars represent a 95% confidence interval.

### 9.1 Prompts

**Abstractive Short-Form Transcript Generation**
*System:*
I will give you a long-form video transcript. Generate a short-form video transcript that can be used for platforms like TikTok, Instagram, and YouTube Shorts. Try to make your generated short-form video transcript in the same style as the long-form video transcript. Write it from the perspective of the main speaker in the long-form video. Use the introduction and conclusion sentence from the long-form video transcript. I will also provide you with a list of visuals that are you not required to but can use to try and add more visuals to your short-form video transcript. Do not include emojis or hashtags in your response or any formatting like "Here are the transcripts:" or "Transcript:". Also give your transcript a title and use the unique separator: <SEP> to separate transcript and title. Since speech will be generated from this transcript please convert written symbols or abbreviations into their full verbal forms. For example you would make "tbsp" into "tablespoon", "1/2" into "one half", "BBQ" into "barbecue".
*User:*
Here is the long-form video transcript and the list of visuals for you to generate a short-form video transcript from.
Here is the list of visuals: **[visual concepts]**
Here is the long-form video transcript: **[long-form transcript]**

**Visual Concepts Extraction**
*System:*
I'm creating a short-form video from a long-form video. I have a sentence that is part of my short-form transcript that I will use to generate speech for my short-form video. I need you to identify the most relevant visual concepts (visually concrete words or phrases) within the sentence that would make good shots for my short-form video. Only recommend visual concepts that viewers would find interesting and relevant. To avoid extra shots, combine similar visual concepts. I will also provide you with previous sentences to help put the sentence into context. Embed your recommended visual concepts directly into the sentence I give you. I will also give you descriptions of the long-form video shots to help you make sure your recommended visual concepts can be found in the long-form video. Use this format to embed a visual concept: exact sentence text <VIS>visual concept</VIS> exact text from sentence. Do not include any statement like "Here is the sentence with embedded visual concepts and descriptions."
*User:*
Previous sentences: **[previous sentences]**
Here is the sentence for you to identify important visual concepts: **[sentence]**
Here are the long-form video shot descriptions to help you make sure your recommended visual concepts can be found in the long-form video: **[long-form shot descriptions]**

**GPT-4o Scoring**
*System:*
I will provide you with N images. You will score these N images from 0 to 1 based on how well they match a visual concept I provide you (for example "spiral staircase"). I will provide this visual concept embedded in a window of its surrounding speech but I will highlight it like this: ***visual concept***. For added context, I will provide for each of the N images a description (describing the image) and its speech. You will return N scores. You will also provide me with a short description for each image. You will return N short descriptions. Do not skip any scores or descriptions for any reason including if you get a black image (you can simply give it a score of 0 and describe it as a black image). Do not provide any statement like "Here are the scores and descriptions" or "Score" with your score. Format your output like this:
Description 1
Score 1
Description 2
Score 2
...
Description N
Score N
*User:*
Here are **[N]** images for you to score and describe from 0 to 1 based on how well they match this visual concept: **[embedded visual concept]**. Please remember if you get a black image to give it a score of 0 and describe as a black image.
Here are the images to score along with their description and speech:
**[images]**
**[image descriptions]**
**[speech (comes from the shot corresponding to each image)]**

**Segmentation**
*System:*
Your task is to take a transcript and split it into segments that are logical and coherent. Each segment should come from the transcript. Do not write any new text or remove any text. Segment the introduction and conclusion with no extra information. Please title each segment. Your output should look like this:
<TITLE> a title
<SEG> a segment

|    | Category  | Duration (Original) | Abstractive+Extractive | Abstractive | Extractive | Source |
|----|-----------|---------------------|------------------------|-------------|------------|--------|
| V1 | Travel    | 13min 2s            | 1min 16s               | 1min 1s     | 1min 2s    | [16]   |
| V2 | Cooking   | 13min 41s           | 1min 25s               | 1min 20s    | 1min 6s    | [17]   |
| V3 | Spotlight | 7min 25s            | 1min 18s               | 53s         | 1min 5s    | [18]   |
| V4 | Travel    | 13min 16s           | 1min 6s                | 50s         | 1min 3s    | [19]   |
| V5 | Travel    | 17min 18s           | 1min 6s                | 1min 18s    | 1min 6s    | [20]   |
| V6 | Home Tour | 15min 12s           | 1min 2s                | 54s         | 1min 5s    | [21]   |

Table 2: Overview of videos used in the results evaluation.

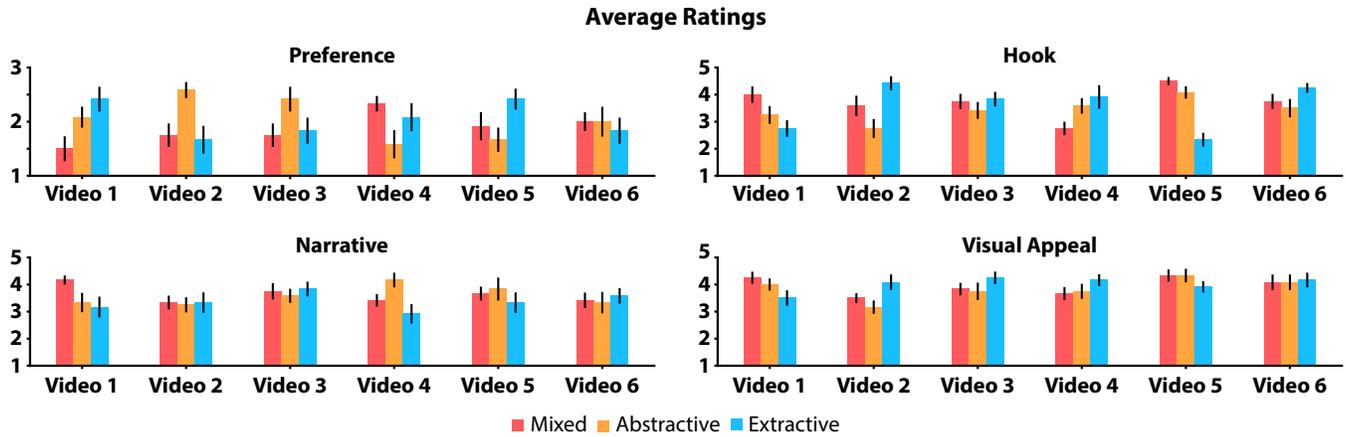

Figure 9: Result evaluation ratings grouped by video for the mixed abstractive-extractive, abstractive, and extractive methods. A shows the average ranking for each method across all 12 participants ranging from 1 to 3 where a lower ranking is better. B, C, and D show the average rating between 1 and 5 for the dimensions *Hook*, *Narrative*, and *Visual Appeal* where higher is better. Error bars represent a 95% confidence interval

<TITLE> a title
<SEG> a segment
<TITLE> a title
<SEG> a segment
<TITLE> a title
<SEG> a segment
*User:*
Here is the transcript that you will be splitting into logical and coherent segments. Each segment should come from the transcript. Do not write any new text or remove any text. Please title each segment. Transcript: **[transcript]**

**Generate Alternative Text**
*System:*
I have a shot in my short-form video that I need its speech to be written. I will provide you with the keyframe of the shot I need speech written for. I will also provide you with the keyframe and speech of the shot before and after this shot for added context. This should help you write speech that fits naturally into my short-form video. It is very important the speech you write fits coherently with the speech before and after it. I will also give you the duration of the shot that you need to write speech for to give you an idea of how long the written speech should be. When writing speech, please do not hallucinate information like names or locations and write casually since this is for social media platforms like TikTok and Instagram. I will provide some extra context to help you. Here is the format I will provide the information in:
CONTEXT: extra context to help you
BEFORE:
Before keyframe
Before speech
REPLACE:
Keyframe of shot to replace
Duration of shot to replace
AFTER:
After keyframe
After speech
Note the before section will be empty if you are replacing the introduction speech, and the after section will be empty if you are replacing the conclusion speech. Treat these replacements like you are replacing an introduction or conclusion. Your output should be in this format:
Before speech<NEW>New written speech</NEW>After speech
*User:*
For my shot that needs speech written here is its keyframe and duration along with the keyframe and speech for the shot before and after:
CONTEXT: **[extra context]**

BEFORE:
**[before keyframe]**
**[before speech]**
REPLACE:
**[replace keyframe]**
**[replace duration]**

AFTER:
**[after keyframe]**
**[after speech]**